\documentclass[trackchanges, twocolumn, twocolappendix]{aastex701}

\usepackage[LGR, T1]{fontenc}

\DeclareTextAccentDefault{\accdasia}{LGR}
\DeclareTextAccentDefault{\acctonos}{LGR}

\begin{document}

\title{The Pristine HeII Emitter near GN-z11: Constraining the Mass Distribution of the First Stars}

\author[orcid=0009-0006-4326-6097]{Elka Rusta}
\affiliation{Dipartimento di Fisica e Astronomia, Università degli Studi di Firenze, Largo E. Fermi 1, 50125, Firenze, Italy}
\affiliation{INAF/Osservatorio Astrofisico di Arcetri, Largo E. Fermi 5, 50125, Firenze, Italy}
\email[show]{elka.rusta@unifi.it}  

\author[orcid=0000-0001-7298-2478]{Stefania Salvadori}
\affiliation{Dipartimento di Fisica e Astronomia, Università degli Studi di Firenze, Largo E. Fermi 1, 50125, Firenze, Italy}
\affiliation{INAF/Osservatorio Astrofisico di Arcetri, Largo E. Fermi 5, 50125, Firenze, Italy}
\email{stefania.salvadori@unifi.it}  

\author[orcid=0000-0002-4985-3819]{Roberto Maiolino}
\affiliation{Kavli Institute for Cosmology, University of Cambridge, Madingley Road, Cambridge CB3 0HA, UK}
\affiliation{Cavendish Laboratory, University of Cambridge, 19 JJ Thomson Avenue, Cambridge CB3 0HE, UK}
\affiliation{Department of Physics and Astronomy, University College London, Gower Street, London WC1E 6BT, UK}
\email{rm665@cam.ac.uk}  

\author[orcid=0000-0001-5487-0392]{Viola Gelli}
\affiliation{Cosmic Dawn Center (DAWN), Denmark}
\affiliation{Niels Bohr Institute, University of Copenhagen, Jagtvej 128, 2200 Copenhagen N, Denmark}
\email{viola.gelli@nbi.ku.dk}   

\author[orcid=0000-0002-3524-7172]{Ioanna Koutsouridou}
\affiliation{Dipartimento di Fisica e Astronomia, Università degli Studi di Firenze, Largo E. Fermi 1, 50125, Firenze, Italy}
\email{ioanna.koutsouridou@unifi.it}  

\author[orcid=0000-0002-6719-380X]{Stefano Carniani}
\affiliation{Scuola Normale Superiore, Piazza dei Cavalieri 7, 56126 Pisa, Italy}
\email{stefano.carniani@sns.it}

\author[orcid=0000-0003-4891-0794]{Hannah {\"U}bler}
\affiliation{Max-Planck-Institut f\"ur extraterrestrische Physik, Gie{\ss}enbachstra{\ss}e 1, 85748 Garching, Germany}
\email{hannah@mpe.mpg.de}

\author[orcid=0000-0002-9889-4238]{Alessandro Marconi}
\affiliation{Dipartimento di Fisica e Astronomia, Università degli Studi di Firenze, Largo E. Fermi 1, 50125, Firenze, Italy}
\affiliation{INAF/Osservatorio Astrofisico di Arcetri, Largo E. Fermi 5, 50125, Firenze, Italy}
\email{alessandro.marconi@unifi.it}  

\author[orcid=0000-0001-7144-7182]{Daniel Schaerer}
\affiliation{Department of Astronomy, University of Geneva, Chemin Pegasi 51, 1290 Versoix, Switzerland}
\email{daniel.schaerer@unige.ch}  

\correspondingauthor{Elka Rusta}

\begin{abstract}
The properties of the first metal-free stars remain largely unknown, and so far, the only data-driven constraints on their mass distribution (IMF) come from near-field cosmology. 
Here, we interpret new observations of the C1 and C2 components of \textit{Hebe}, the HeII emitter near the galaxy GN-z11.
Using a locally calibrated model, we robustly confirm the pristine (PopIII) nature of both components, showing that the measured upper limits on metal lines can only be reproduced by galaxies with $>50\%$ of their stellar mass in PopIII stars. We find that C1 is consistent with a purely PopIII system and adopt a simple parametric approach to infer the implications for the PopIII IMF and stellar mass. The observed $\rm HeII/H\gamma$ ratio excludes steep IMFs, favoring top-heavy distributions, especially for young stellar ages ($\leq 1$ Myr). Combined with the HeII luminosity, this implies a total PopIII stellar mass of $2 \cdot 10^4 < M_\star/M_\odot < 6 \cdot 10^5$. While degeneracies between IMF, stellar mass, and age remain, adopting the lower stellar masses predicted by simulations ($M_\star < 10^5\,M_\odot$) strengthens the preference for top-heavy IMFs. Combining these results with near-field constraints, which instead exclude the flattest IMFs, we define a data-driven range of viable PopIII IMFs, linking characteristic mass and slope.
This work demonstrates that direct observations of high-$z$ PopIII systems can place independent constraints on the IMF of the first stars, opening a new window on their formation and properties.

\end{abstract}


\section{Introduction} 
\label{sec:intro}
Uncovering the nature of the first stars is one of the key goals of present-day cosmology and of the JWST mission. Also known as Population III (hereafter PopIII) stars, they produce
the first ionizing photons, the first elements heavier than helium, and the first stellar black holes (BHs), thus profoundly affecting the galaxy formation process \citep[e.g.,][]{Bromm2013}. In particular, the mass distribution of PopIII stars, i.e. the {\it PopIII IMF}, regulates the number and type of photons and chemical elements, and BHs produced in the early Universe, thereby determining how reionization and metal enrichment began, while their role in supermassive BH seeding remains uncertain.

However, the PopIII IMF remains largely unknown. Cosmological simulations of the first star-forming systems predict different typical masses for PopIII stars \citep[e.g.,][]{klessen+23}. Thus, so far, the only constraints on the PopIII IMF come from near-field cosmology. The non-detection of long-lived metal-free stars in ultra-faint dwarf galaxies sets a lower limit on the peak of the PopIII IMF, $M_{ch}> 1M_{\odot}$ \citep{rossi2021ultra}. In the Galactic halo, the absence of stars showing the key chemical signatures of Pair Instability Supernovae (PISNe) \citep{salvadori2019probing} excludes the flattest PopIII IMFs \citep{Koustouridou2024}.

Ultimately, PopIII stars were likely more massive than subsequent generations of metal-enriched (PopII) stars. However, a wide variety of plausible PopIII IMFs remains: the peak can range from one to hundreds of solar masses, and the slope can be either flatter or steeper than Salpeter, although not completely flat. Can we use \textit{JWST} observations of high-$z$ galaxies to further constrain the PopIII IMF?

The highly ionizing emission of PopIII stars produces characteristic Helium II (HeII) recombination lines, like $\rm HeII\lambda1640$ \citep[e.g.][]{tumlinson+00, peng01, zackrisson+11, inoue+11, Nakajima+22, lecroq+25}. Hence, we expect a strong HeII in the emission of systems hosting PopIII stars, combined with signatures of either metal-free or metal-poor gas enriched by the first PopIII supernovae.

Despite extensive searches, only a handful of PopIII candidates have been identified with \textit{JWST} so far. 
All but one are at least {\it very metal-poor systems} at $z<7$ \citep[e.g.][]{vanzella+23, Nakajima+25}, with some potentially being metal-free \citep{morishita+25, vanzella+26}, yet they currently lack the HeII detection expected for PopIII stars.  
Only the $z=10.6$ companion of GN-z11 matches the expected properties of a pure PopIII galaxy, showing a combination of both a strong HeII emission, with $\rm EW(HeII\lambda1640) > 20\,
$\AA, and absence of detectable metal lines \citep{maiolino+24}. 

Deeper \textit{JWST} observations reported in two companion papers have now revealed that this system, which we will dub \textit{Hebe}\footnote[1]{HElium Balmer Emitter. In ancient Greek mythology Hebe (\accdasia{}\acctonos{}H$\beta \eta$) is the goddess of youth, daughter of Zeus and Hera.}, consists of two components, C1 and C2, both showing HeII emission \citep[][]{maiolino+26}, along with tight upper limits on rest-frame UV and optical metal lines, and an H$\gamma$ detection in C2 \citep[][]{ubler+26}.

In this Letter, we use the model presented in \citealt{rusta+25} to interpret the new observations of \textit{Hebe} and assess their implications, addressing urgent questions: Is the companion of GN-z11 a truly pristine PopIII system, or is it also consistent with a very metal-poor PopII (or hybrid) galaxy? Can we use the observed spectral features of its components to further constrain the PopIII IMF?

\section{Methods: recap of the model}
\label{sec:methods}
We adopt the model of \citealt{rusta+25}, which combines the locally calibrated galaxy formation code {\tt NEFERTITI}, stellar population synthesis libraries, and the publicly available code {\tt CLOUDY}. 
With this modeling framework, we can self-consistently trace the chemical enrichment and spectral evolution of PopIII galaxies, and thus combine constraints on the PopIII IMF from both near- and far-field cosmology.

As defined in \citealt{rusta+25}, we subdivide PopIII galaxies into: \textit{pristine} - only PopIII stars and metal-free gas, \textit{self-polluted} - only PopIII stars and PopIII-enriched gas, \textit{PopIII-rich/mid/poor hybrids} - mixture of PopIII and PopII stars, respectively with $>50\%$, $25-50\%$, and $<25\%$ of the total stellar mass in metal-free stars.

\subsection{The {\tt NEFERTITI} Semi-Analytical Model}

We employ the {\tt NEFERTITI} model coupled with a dark matter cosmological simulation of a Milky Way analog \citep[see][]{Koutsouridou2023} to simulate PopIII galaxies at high-$z$ \citep[][]{rusta+24}. Given its calibration with near-field cosmology data \citep{Koutsouridou2023, Koustouridou2024, Koutsouridou2025}, the model is particularly suitable for studying the properties of the first stars. {\tt NEFERTITI} follows the formation and evolution of individual stars, starting from pristine gas, and self-consistently traces the chemical enrichment of the gas. Stars are formed following a Larson-type IMF \citep[][]{larson1998early}:
\begin{equation}
\label{e:Larson}
\phi(m_\star) = \frac{d N}{d m_\star} \propto m_\star^{-x} {\rm exp} \bigg( - \frac{M_{\rm ch}}{m_\star} \bigg),
\end{equation}
where $x$ represents the slope of the IMF and $M_{\rm ch}$ the characteristic mass, i.e., the IMF peak. For PopIII stars we assume a mass range $m_\star = [0.8 - 1000]\: {\rm M_\odot}$, a fiducial value of $x=2.35$, and explore $M_{\rm ch}=[1, 10, 70]\:{\rm M_\odot}$ \citep[see][]{Koutsouridou2023,Koustouridou2024}. 
For PopII stars, a mass range $m_\star = [0.08 - 100]\: {\rm M_\odot}$, $x=2.35$ and $M_{\rm ch} = 0.35\: {\rm M_\odot}$.

For this project, we analyze a set of 250 PopIII galaxies and 205 metal-poor PopII galaxies (with $\rm log\,O/H +12 < 7.5$) simulated with {\tt NEFERTITI} at $z=10.6$, the observed redshift of \textit{Hebe}. 

\subsection{The Spectral Synthesis and {\tt CLOUDY} Models}

We produce the synthetic spectra of our set of {\tt NEFERTITI} galaxies by constructing the stellar continuum and then running photoionization models with {\tt CLOUDY} \citep[C23 release][]{chatzikos+23, ferland+98} to obtain the nebular emission. For the stellar emission, we use individual stellar spectra and evolutionary tracks from \citealt{Schaerer2002} for PopIII stars, both to account for the incomplete IMF sampling and to freely vary the IMF shape and range, and IMF-integrated spectra from \citealt{schaerer+03} and \citealt{zackrisson+11} for PopII stars. We then input both the total stellar continuum and the chemical composition of the gas into the {\tt CLOUDY} code to self-consistently trace the metal enrichment in the transition from PopIII to hybrid galaxies.

We adopt the same {\tt CLOUDY} modeling assumptions as in \citealt{rusta+25},
with gas density $n_H=10^3 \rm cm^{-3}$ and ionization parameters $\rm logU = [0, -0.5, -1, -2]$.
We also run a high-density case with $n_H=10^6 \rm cm^{-3}$ and $\rm logU = -2$.
Considering the different PopIII galaxy stages and parameters explored, we ultimately obtain a set of 5000 photoionization models for PopIII galaxies, and 820 for metal-poor PopII galaxies. 

\begin{figure}[t]
    \includegraphics[width=1.\hsize]{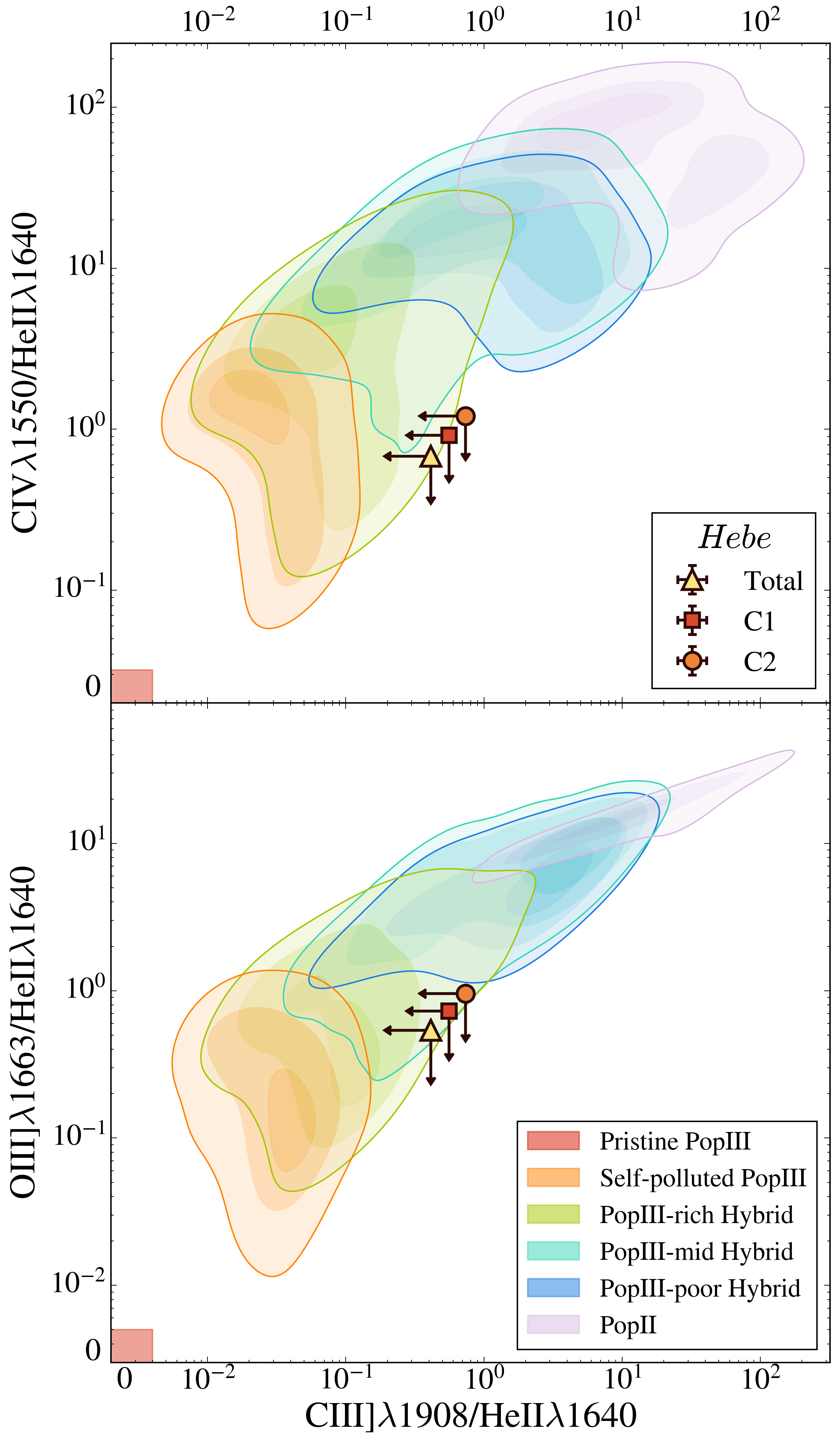} 
    \hspace{0.01\hsize}
    \caption{Density distributions of our {\tt NEFERTITI} models at $z=10.6$ for PopIII galaxies at different evolutionary stages (red: {\it pristine}; orange: {\it self-polluted}; green: {\it PopIII-rich hybrids}) and for metal-poor PopII galaxies ({\it purple}). The solid contours include $68\%$ of the galaxy population, for $\rm logU = [-2, -1, -0.5, 0]$ altogether. The errorbars are the $3\sigma$ observational upper limits presented in \citealt{maiolino+26}.}
  \label{fig:fig1}
\end{figure}

\label{sec:pristine}

\begin{figure*}[t]
    \centering
    \includegraphics[width=0.65\hsize]{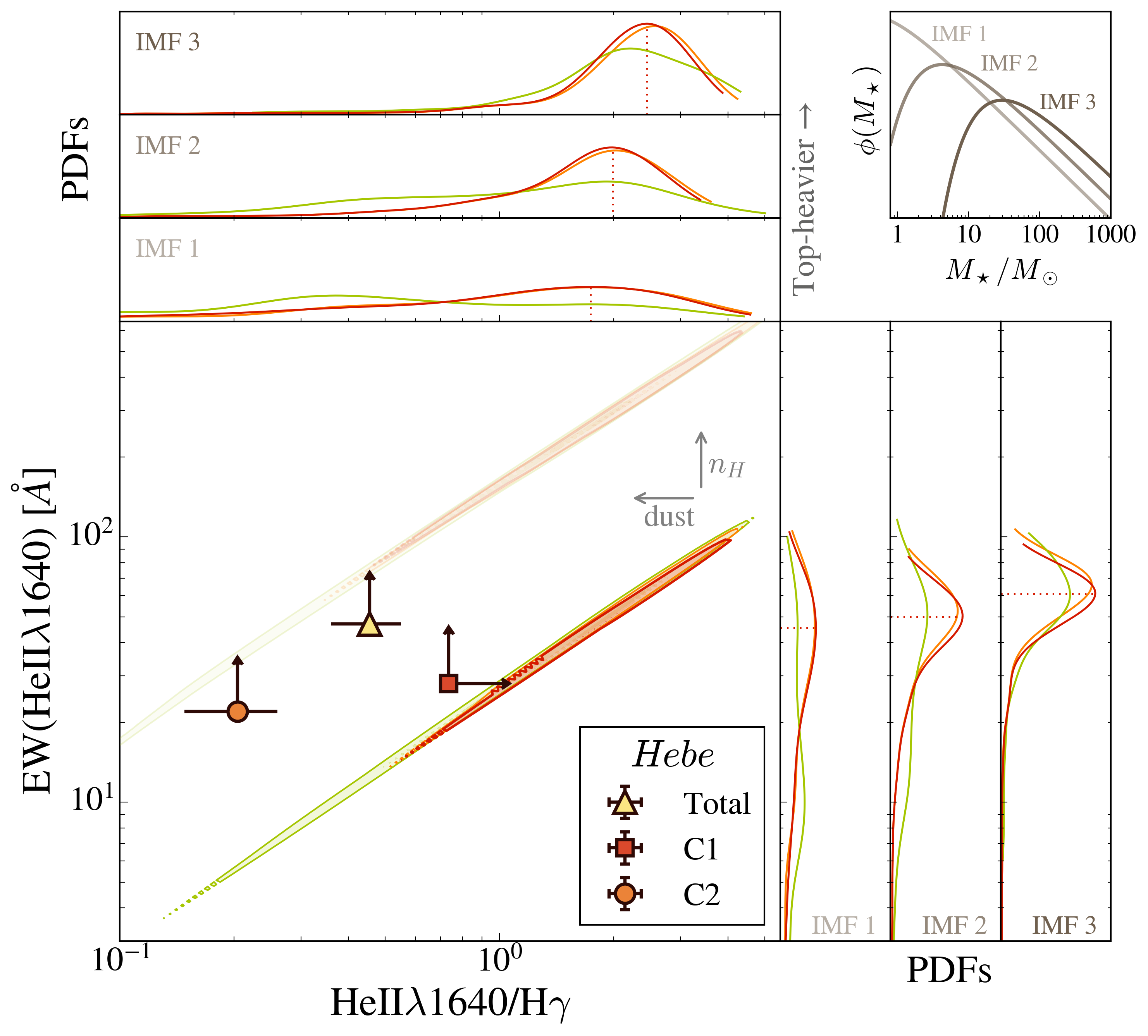} 
    \hspace{0.01\hsize}
    \caption{EW(HeII) vs $\rm HeII/H\gamma$ for the sub-sample of {\tt NEFERTITI} galaxies at $z=10.6$ that match the upper limits on metal emission lines (\ref{fig:fig1}), shown for three PopIII IMFs as indicated in the top-right scheme (Salpeter slope $x=2.35$, peak mass $M_{ch} = [1, 10, 70] M_\odot$). For each IMF, we show the PDFs of different PopIII galaxy types: {\it pure} (red), {\it self-polluted} (orange) and {\it PopIII-rich hybrids} (green). Arrows indicate how dust and/or higher gas densities would modify the observed properties of these PopIII galaxies. We also report in lighter colors the high-density models (i.e. $n_H = 10^6 cm^{-3}$). The errorbars are the $3\sigma$ data from \citealt{maiolino+26} and \citealt{ubler+26}.
 }
  \label{fig:fig2}
\end{figure*}

\section{The pristine nature of $\rm Hebe$}
\label{sec:results}

\begin{figure*}[t]
    \includegraphics[width=0.495\hsize]{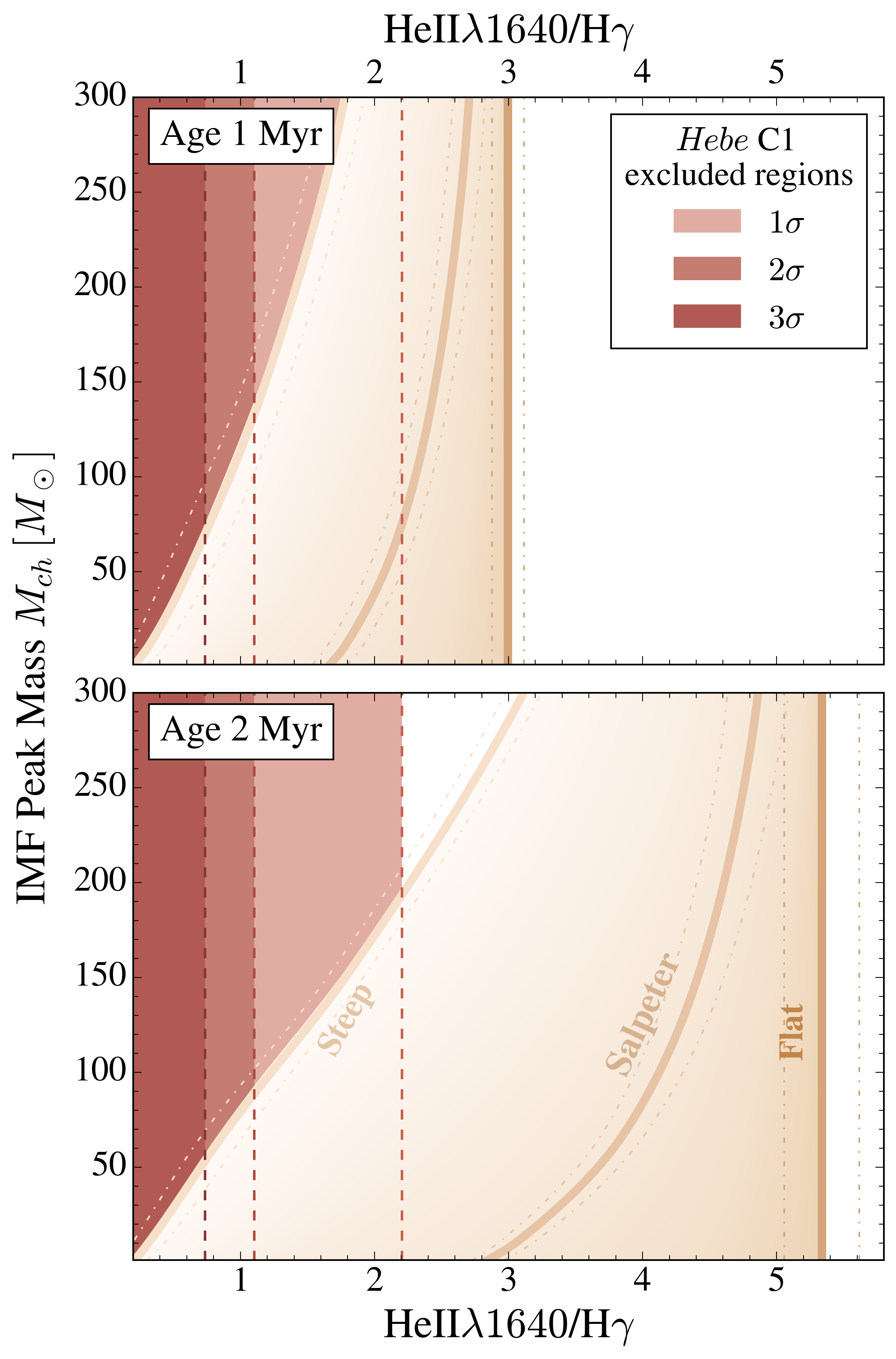} 
    \hspace{0.01\hsize}
    \includegraphics[width=0.5\hsize]{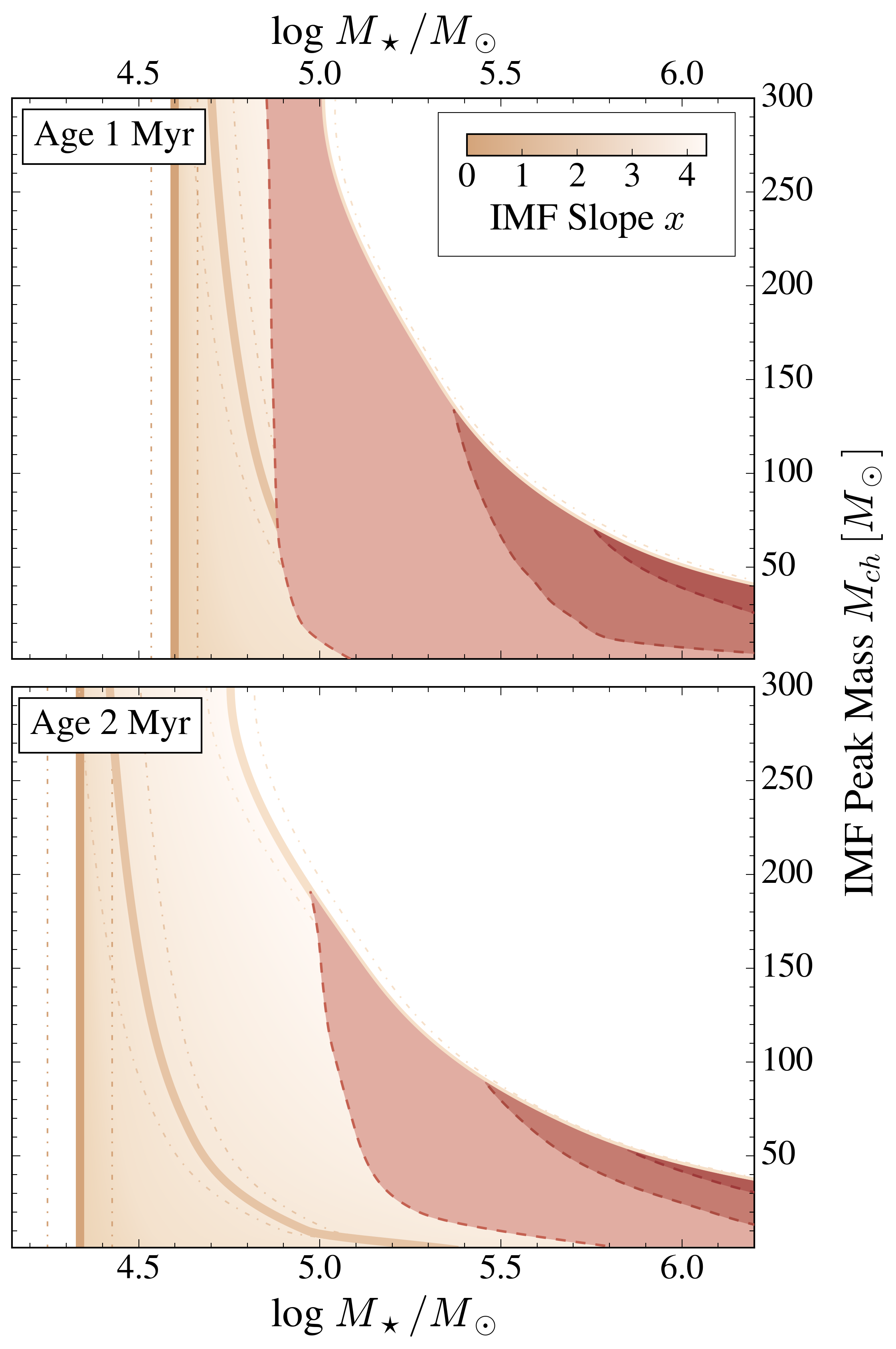}
    \caption{{\it Left}: predicted $\rm HeII/H\gamma$ emission from PopIII stars for ages 1~Myr ({\it top}) and 2~Myr ({\it bottom}), for different $M_{ch}$ and IMF slope ({\it beige regions}). The red areas indicate the $3\sigma$, $2\sigma$ and $1\sigma$ lower limits measured by \citealt{maiolino+26} in the C1 component of \textit{Hebe}. 
    {\it Right}: mass of PopIII stars required to reproduce the observed $\rm L_{HeII}$ value of C1 for ages 1~Myr ({\it top}) and 2~Myr ({\it top}), using different $M_{ch}$ and IMF slopes. Red areas mark IMFs excluded by the $\rm HeII/H\gamma$ constraints (left panels). 
    In {\it both left and right panels} the beige scale shows different PopIII IMF slopes (see legend in the top-right panel) with solid lines highlighting the flat ($x=0$), Salpeter, and steep ($x=4.35$) slopes. For these three examples, the dotted lines represent the scatter due to $\rm logU=[0,-2]$, and the error of the measured $L_{HeII}$ value.} 
  \label{fig:fig3}
\end{figure*}

We first explore the nature of {\it Hebe} by using our PopIII models to interpret the observed line ratios reported by \citealt{maiolino+26} for the individual C1 and C2 components. Fig.~\ref{fig:fig1} compares the measured 3$\sigma$ upper limits on the CIV, CIII], and OIII] emission lines\footnote[2]{Notation: CIV for CIV$\lambda\lambda 1548, 1551$, CIII] for CIII]$\lambda1907$ + $\lambda1909$, OIII] for OIII] $\lambda \lambda 1661, 1666$, HeII for HeII$\lambda1640$.} relative to HeII, with the predictions of our model for PopIII galaxies in different evolutionary stages, and for metal-poor PopII galaxies (see labels).

We clearly see that even upper limits on these metal species are highly informative when compared with model predictions. Indeed, we find that both C1 and C2 are not consistent with either PopII galaxies or PopIII-poor hybrids, and are instead fully consistent with being {\it at least} PopIII-rich galaxies, i.e., with $> 50\%$ of their stellar mass in metal-free stars, and with having $\rm log\,O/H < 7.4$.
Moreover, they are potentially consistent with pristine (or self-polluted) PopIII galaxies, hence composed exclusively of PopIII stars and containing metal-free (or PopIII-enriched) gas, with C1 appearing more pristine than C2. 
This PopIII scenario is further supported by the two companion papers \citep[][]{maiolino+26, ubler+26}, showing that this is the most plausible among others (e.g. direct collapse black holes). 

Based on this result, we now focus on models with $>50\%$ of stellar mass in PopIII stars and move on to explore the impact of the PopIII IMF on the HeII emission.
In Fig. \ref{fig:fig2} we show the EW(HeII) versus the $\rm HeII/H\gamma$ ratio predicted for pristine PopIII, self-polluted PopIII, and PopIII-rich hybrids, together with {\it Hebe} data. 
We note that with increasing characteristic mass, i.e., for progressively top-heavier PopIII IMFs, the distribution of PopIII galaxies peaks at higher EW(HeII) and $\rm HeII/H\gamma$ values, and the peak also becomes more pronounced. Indeed, for top-heavy PopIII IMFs there is a higher probability of forming very massive stars, which have the strongest $\rm HeII$ emission. 
More specifically, for increasing $M_{ch}$ we find that the probability P($\rm EW(HeII) > 50$ \AA)\, goes from $26\%$ to $40\%$, and reaches $69\%$ for the most top-heavy IMF. 

Interestingly, C1 is fully consistent with the peaks of the distribution of our PopIII galaxy models for all three IMFs explored, and could potentially lie on the EW(HeII) versus $\rm HeII/H\gamma$ relation predicted for PopIII stars. 

Conversely, C2 shows a lower $\rm HeII/H\gamma$ ratio while maintaining the high EW(HeII) characteristic of PopIII stars. 
However, unlike the diagnostics shown in Fig.~\ref{fig:fig1}, the $\rm HeII/H\gamma$ ratio may be significantly affected by dust as it combines UV and optical lines. 
The presence of dust would decrease the $\rm HeII/H\gamma$ without affecting EW(HeII), implying that C2 could be a self-polluted or hybrid PopIII system enriched 
by recent PopIII SNe. 
As discussed in \citealt{maiolino+26}, an extinction of only $A_V \sim 0.25$ would be sufficient to match our PopIII predictions when adopting the extinction curve from \citealt{Sun2026}. Using instead the flatter relation derived by \citealt{Markov2025} we estimate $A_V \sim 1.4$. This implies larger dust masses $\leq 50\,M_\odot$ for a very compact configuration, which can however be provided by a few Pair Instability Supernovae, each one producing $>10\,M_{\odot}$ of dust \citep{Nozawa2003, Schneider2004}. 
An alternative scenario involves very high gas densities, as shown by the high-density ($n_H=10^6 cm^{-3}$) models reported in Fig. \ref{fig:fig2} with lighter colors. For PopIII stars, the continuum at the wavelength of HeII is dominated by 2-photon emission \citep[e.g.][]{Raiter+10}, which is heavily suppressed at $n_H \geq 10^5\rm cm^{-3}$. For this reason, the high-density models have much higher EW(HeII) while maintaining a similar $\rm HeII/H\gamma$ ratio.
This would also suggest that C2 is a hybrid galaxy, consistent with the interpretation of \citealt{ubler+26}.

Given these findings, we will now focus on the interpretation of the C1 component of {\it Hebe} as an archetypal pure PopIII system.

\section{Exploring PopIII IMF and stellar mass}
\label{sec:IMF}

\begin{figure}[t]
    \includegraphics[width=1.\hsize]{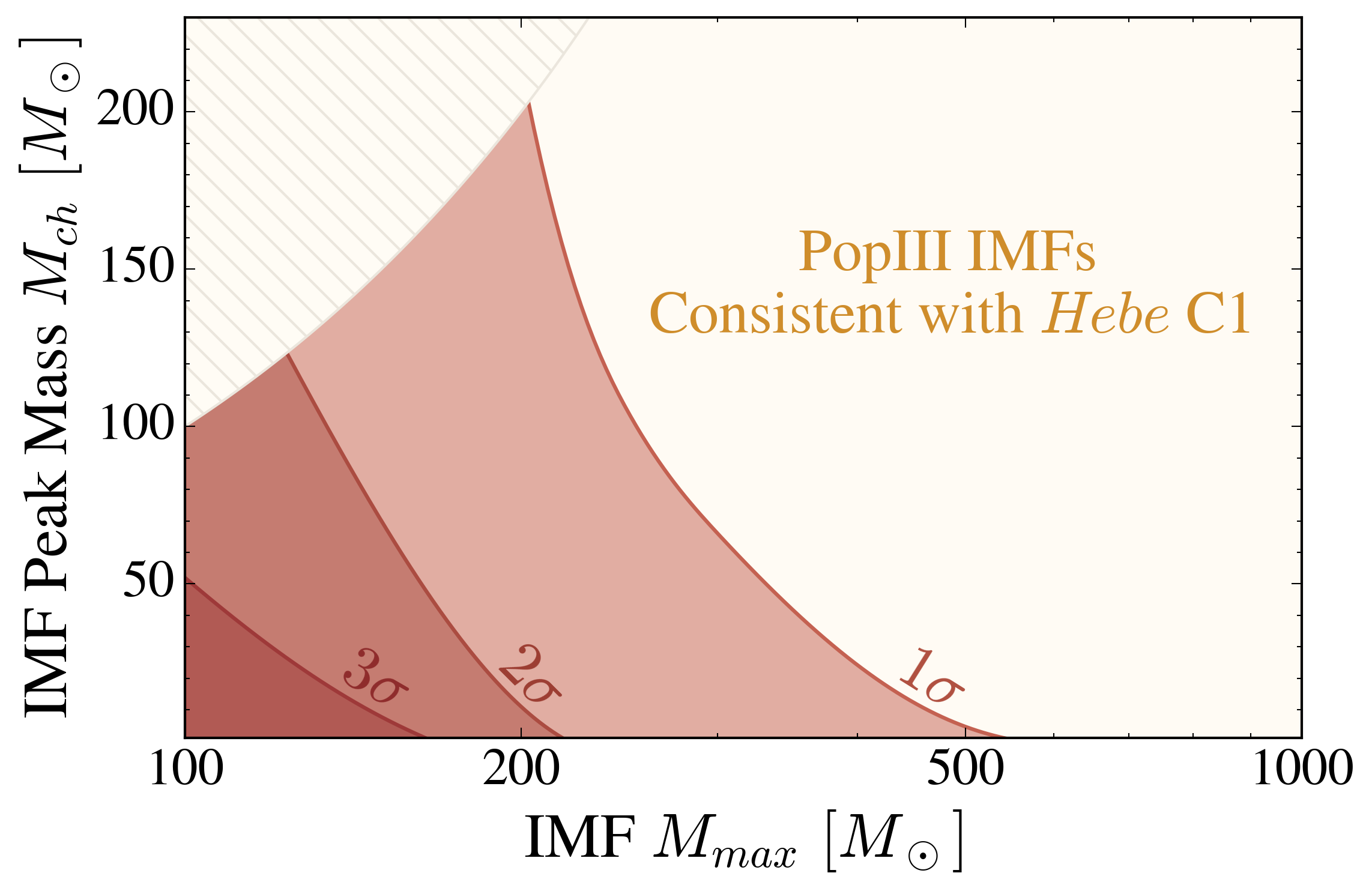} 
    \caption{Constraints on the peak ($M_{ch}$) and maximum mass ($M_{max}$) of the PopIII IMF assuming a Salpeter slope, based on the 1, 2, and 3$\sigma$ upper limits of $\rm HeII/H\gamma$ in the C1 component of \textit{Hebe} (red shaded regions). To maintain the Larson shape of the IMF, the regions with $M_{ch} > M_{max}$ are canceled out. } 
  \label{fig:fig3.2}
\end{figure} 

The system {\it Hebe} lies in a dense and peculiar environment,  being the companion of GN-z11, one of the most massive galaxies observed at high redshift. In contrast, our data-calibrated cosmological model {\tt NEFERTITI} focuses on the assembly of the Milky Way, which corresponds to a smaller fluctuation of the density field. As a result, at $z=10.6$ the model predicts PopIII galaxies that are less luminous than the GN-z11 companion, with total stellar masses $M_\star \leq 4 \cdot10^{3}\,M_{\odot}$ (see Fig. \ref{fig:fig5} in the Appendix). While this does not affect comparisons of emission line ratios or EWs, which are independent of the total stellar mass, it prevents a direct interpretation of the observed line luminosities.

Here we develop a simple parametric approach to use the observed $\rm HeII/H\gamma$ ratio and the HeII luminosity, $L_{\rm HeII} = (5.1 \pm0.9) \cdot 10^{40} \rm erg/s$ of C1 to constrain the PopIII IMF and stellar mass.
To this end, we compute the synthetic emission from a single PopIII star formation episode at different ages (1 Myr steps), using the same {\tt CLOUDY} parameters as in Sec.~\ref{sec:methods}, and with a fully sampled IMF. This assumption is discussed in Appendix \ref{app:a}, where Fig. \ref{fig:fig5} shows how the predicted HeII luminosity varies with the total mass of PopIII stars formed, $M_\star$, when the fiducial IMF is randomly or fully sampled. To explore the IMF widely, we vary both the peak $M_{ch}$ in the range $[1-300]\, M_\odot$, and the slope $x$ between $0$ (flat) and $4.35$ (steep), keeping a fixed mass range of $[0.8-1000] M_\odot$ (see equation \ref{e:Larson}). The validity of this assumption is discussed at the end of the section (see Fig. \ref{fig:fig3.2}).
We focus on the emission at 1 Myr and 2 Myr after the star-formation episode to provide a reasonable range of results. Indeed, immediately after the birth of PopIII stars (0~Myrs), the emission takes intermediate values between those considered here, while by 3 Myr the HeII emission is already too faint due to the death of the most massive stars \citep[][]{Schaerer2002}.

As seen also in Fig. \ref{fig:fig2}, $\rm HeII/H\gamma$ depends on the PopIII IMF, but not on the total stellar mass formed, $M_\star$, since it is an emission line ratio. In contrast, line luminosities scale strongly with $M_\star$ (see Fig.~\ref{fig:fig5} for a fixed IMF), motivating our parametric study, shown in Fig.~\ref{fig:fig3}. We thus use $\rm HeII/H\gamma$ to exclude PopIII IMFs which are not consistent with the observed ratio, independently of $M_\star$ (left panels). For the remaining PopIII IMFs, we then determine the stellar mass required to obtain the observed $L_{\rm HeII}$ (right panels). 
In both cases we show the results of our parametric study for ages 1 Myr (first row) and 2 Myr (second row).

At an age of 2 Myr, we obtain the highest HeII luminosities and $\rm HeII/H\gamma$ ratios, corresponding to the peak of the HeII emission produced by the most massive PopIII stars in the adopted stellar population libraries \citep[][]{Schaerer2002}. 
The left panels of Fig. \ref{fig:fig3} show the regions excluded by the lower limits on the observed $\rm HeII/H\gamma$ of C1 at different confidence levels. Steep PopIII IMFs with low $M_{ch}$ fail to reproduce the high observed value ($\rm HeII/H\gamma > 0.7$ at $3\sigma$), and can therefore be ruled out. 
However, the minimum $M_{ch}$ implied by the $3\sigma, 2\sigma$, and $1\sigma $ limits depends on both the assumed stellar age and IMF slope. For example, for an age of 1 Myr and a Salpeter slope ($x=2.35$) the 3$\sigma$ limit requires $M_{\rm ch} \gtrsim 75\,M_\odot$, favoring relatively top-heavy PopIII IMFs. 

The right panels show the stellar masses required for each PopIII IMF to reproduce the HeII luminosity of C1, with the IMFs excluded by the $\mathrm{HeII}/\mathrm{H}_\gamma$ constraints (left panels) indicated in red. 
We find that models matching both the observed $\mathrm{HeII}/\mathrm{H}_\gamma$ ratio and the HeII luminosity predict total stellar masses of 
$10^{4.25 - 5.8} \,M_\odot$, 
with the exact value depending on the stellar population age and the assumed IMF.
The minimum mass is obtained for an age of 2~Myr and a flat PopIII IMF. Conversely, steeper PopIII IMFs require either a larger $M_{\rm ch}$ or a higher stellar mass to reproduce the high observed $L_{\rm HeII}$. We stress that the PopIII IMFs implying the largest stellar masses are already excluded (see left panels of Fig. \ref{fig:fig3}), as they fail to reproduce the observed $\mathrm{HeII}/\mathrm{H}_\gamma$ ratio. 

Additionally, we investigate the impact of varying the PopIII IMF mass range to validate our assumption of $[0.8-1000] M_\odot$. While the minimum mass has a negligible effect on our results, the maximum mass strongly affects our predictions. For this reason, we explore $M_{max} = [100-1000] M_\odot$ for a fixed Salpeter slope. In Fig. \ref{fig:fig3.2} we report the $M_{max}$ which can be excluded, as a function of $M_{ch}$, based on the $\rm HeII/H\gamma$ observed in the C1 component of \textit{Hebe}. These are the most conservative constraints, taken at the age of the peak HeII emission.
We find that a standard $M_{max}=100 M_{\odot}$ can be excluded at $2\sigma$ confidence level if $M_{ch} \in (50-100) M_{\odot}$ and at $3\sigma$ for $M_{ch} \in (1-50) M_{\odot}$.  
For fixed $M_{ch} = 10 M_\odot$, the maximum mass must be at least $150 M_\odot$ ($3\sigma$), $200 M_\odot$ ($2\sigma$), or $460 M_\odot$ ($1\sigma$). Notably, we also find that for a flat PopIII IMF $M_{max} > 200 M_\odot$ at $1\sigma$. Hence, our assumption of a mass range up to $1000 M_\odot$ is in agreement with \textit{Hebe} for all the $M_{ch}$ explored.

Ultimately, our analysis of the C1 component of {\it Hebe} favors a top-heavy PopIII IMF.

\section{Discussion and Conclusions}
\label{sec: conclusions}

The first confirmation of a pristine HeII emitter at $z=10.6$, i.e., {\it Hebe}, showcases the enormous advances made by JWST in the quest for PopIII stars. Direct detections of high-$z$ galaxies hosting PopIII stars are crucial to complement our limited knowledge of the first stars’ properties, currently based on near-field cosmology studies. In this Letter, we use a locally calibrated model to robustly confirm the PopIII nature of {\it Hebe} and explore its implications by combining near- and far-field approaches.

Initially, we investigate the upper limits on UV metal lines, comparing observations with PopIII and PopII galaxy predictions obtained from the {\tt NEFERTITI} model. Adopting a cosmological model that reproduces present-day observations of ancient metal-poor stars allows us to obtain reliable predictions for the early chemical enrichment.
For both the C1 and C2 components in {\it Hebe}, models with $<50\%$ PopIII stellar mass fail to reproduce the observations. We find that C1 is fully consistent with a pure PopIII system, composed exclusively of PopIII stars and surrounded by either pristine gas or gas self-enriched by PopIII supernovae (see Sec.~\ref{sec:methods}). This makes C1 the most compelling candidate for a genuinely pristine stellar system identified to date.

Treating C1 as an archetypal pristine PopIII system, we explore the PopIII IMF, to constrain its shape and characteristic mass, using a Larson-type with $m_\star = [0.8, 1000] M_\odot$ (see equation \ref{e:Larson}). The observed $\rm HeII/H\gamma$ ratio excludes the steepest IMFs, with particularly strong constraints for a very young system, with stellar ages $\leq 1$ Myr. In this regime, IMFs with Salpeter-like slopes and with characteristic masses $M_{\rm ch} < 75M_\odot$ are disfavored at the $1\sigma$ level. 
Since our analysis relies on H and He lines, these IMF constraints remain valid even if C1 is a self-polluted PopIII system: the presence of metals or dust would increase the intrinsic $\rm HeII/H\gamma$, reinforcing our conclusions. Furthermore, in our high-density models ($n_H = 10^6 cm^{-3}$), the $\rm HeII/H\gamma$ ratios are mildly affected, thus maintaining our results valid.

Given these IMF constraints, we infer a total PopIII stellar mass of $ 2\cdot 10^4  < M_\star /M_\odot < 6 \cdot 10^5$ to reproduce the observed HeII luminosity. Our mass estimates are roughly consistent with predictions from cosmological models and simulations at $z\approx 11$ \citep[e.g.,][]{Jaacks2019, ventura2024, Hazlett2025, jeong2026}, although on the higher side, which is expected due to the vicinity to a massive galaxy like GN-z11. Assuming a lower maximum mass for PopIII stars would imply even higher stellar masses, in disagreement with theoretical studies. This motivates our choice of employing a PopIII IMF up to $1000 M_\odot$ and further supports our top-heavy IMF scenario.
However, strong degeneracies remain between IMF, stellar mass, and age (Fig.~\ref{fig:fig3}).

\begin{figure}[t]
    \includegraphics[width=1.\hsize]{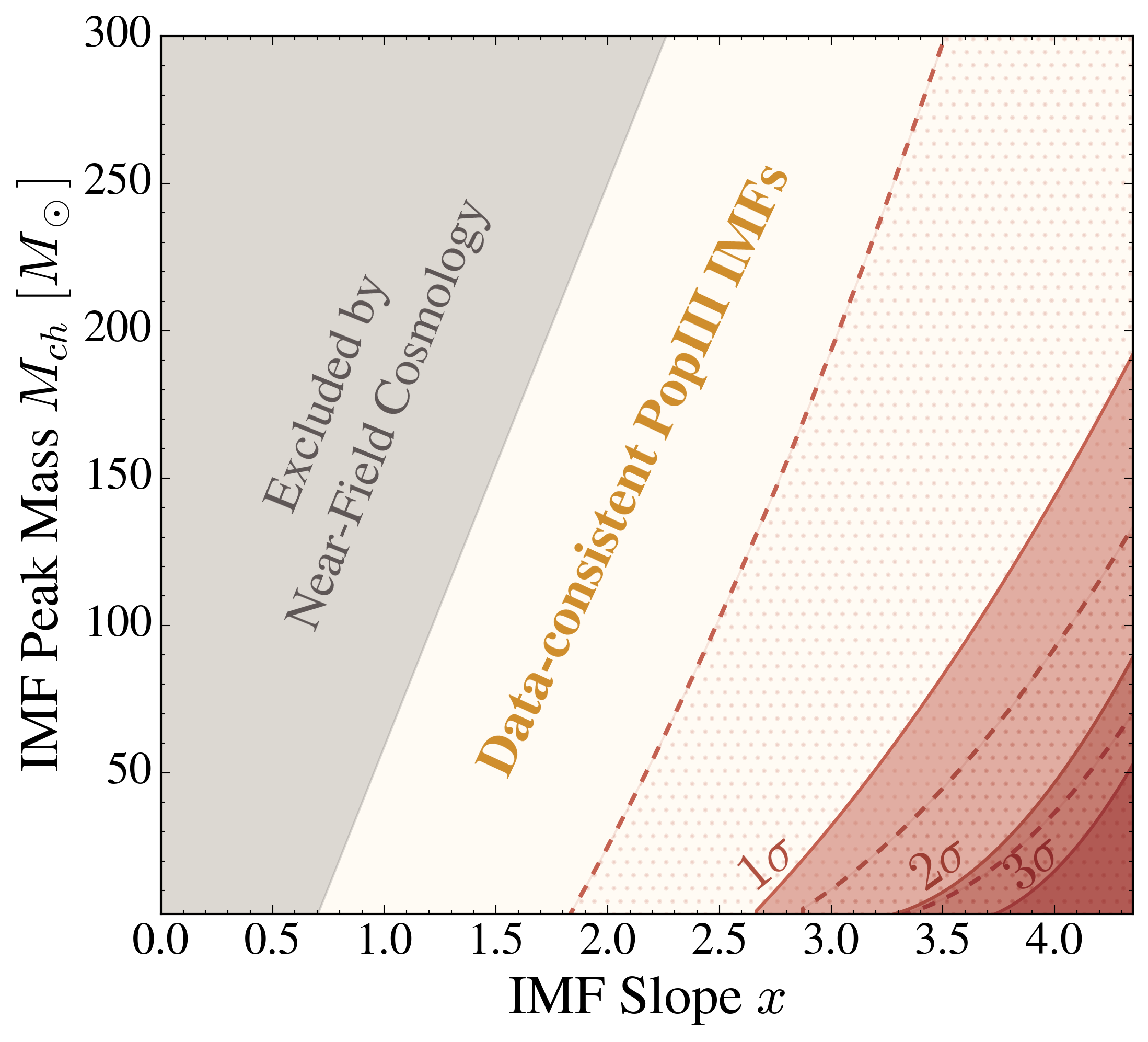} 
    \caption{ Constraints on PopIII IMF shapes combining our results with near-field cosmology studies \citep[gray area, ][]{Koustouridou2024}. Red shaded regions show IMF peaks ($M_{ch}$) and slope ($x$) excluded by this work based on the 1, 2, and 3$\sigma$ upper limits of $\rm He\, II/H\gamma$ in the C1 component of {\it Hebe} (left panels in Fig.~\ref{fig:fig3}). Dotted regions are valid only for ages of 1 Myr.}
   \label{fig:fig4}
\end{figure} 

Extremely top-heavy IMFs, either flat or with a Salpeter slope but a peak at $M_{ch} \geq 140 M_{\odot}$, i.e., in the mass range of PISNe, are predicted for low stellar masses $\approx (2-3)\cdot 10^{4}M_{\odot}$. This is particularly intriguing in light of recent results of a putative high-$z$ PISN detection at $z\approx 15$ \citep[][]{gandolfi+2026, ferrara+2026}, and the prospect of detecting their chemical signatures in high-$z$ gas clouds \citep{Vanni2024}. In this context, C1 may provide direct constraints on the progenitor population of such energetic explosions.

Conversely, steeper PopIII IMFs, such as a Salpeter slope with $M_{ch} = 1M_{\odot}$ for ages of 2~Myr, would require stellar masses $\geq 10^{5}M_{\odot}$. Such high masses cannot be excluded, as {\it Hebe} lies near the luminous galaxy GN-z11 \citep{Venditti2023}, whose Lyman–Werner radiation may suppress H$_2$ cooling in nearby minihalos. However, these masses are higher than typically predicted by simulations that resolve minihalos 
\citep[e.g.,][]{Storck2025}. Models with $M_\star < 10^{5}\,M_\odot$ also imply a top-heavy PopIII IMF, as they exclude all IMFs steeper than a Salpeter slope with $M_{\rm ch} < 10\,M_\odot$ (Fig.~\ref{fig:fig3}). The resulting star formation rates are $(0.018-0.1) M_{\odot}/yr$.  Thus, if PopIII stellar masses are limited to the lower values predicted by simulations, our results effectively require top-heavy IMFs.

Fig.~\ref{fig:fig4} combines our far-field cosmology PopIII IMF constraints with those from near-field Cosmology studies. The {\it direct detection} of a pure PopIII system, C1 in {\it Hebe}, allows us to exclude the least top-heavy Larson PopIII IMFs: $M_{ch} < ax-300$, where 
x is the IMF slope and $a=(111,88,80)$ for 1, 2, and 3$\sigma$ lower limits on  $\mathrm{HeII}/\mathrm{H}_\gamma$. Conversely, {\it indirect studies} of the chemical signature of PopIII supernovae in Galactic halo stars rule out the flattest ones \citep{Koustouridou2024}. 
Together, these complementary approaches bracket the PopIII IMF from both ends, providing
a data-driven range of viable PopIII IMFs. The most likely IMFs satisfy these conditions: $176x - 331 \leq M_{\rm ch} \leq 191x - 132$, implying a narrower stellar mass range for C1 in {\it Hebe}, $ 2.5 \cdot 10^4 \lesssim M_\star \lesssim 10^5$. A precise measurement of $\rm HeII/H\gamma$ would define a direct relation between $M_{\rm ch}$ and the IMF slope, motivating deeper observations of PopIII candidates.
Future JWST observations will therefore be critical to transform these constraints into a direct measurement of the PopIII IMF.

\begin{acknowledgments}
This project received funding from the ERC Starting Grant NEFERTITI H2020/804240 (PI: Salvadori).
H\"U thanks the Max Planck Society for support through the Lise Meitner Excellence Program. H\"U acknowledges funding by the European Union (ERC APEX, 101164796). Views and opinions expressed are however those of the authors only and do not necessarily reflect those of the European Union or the European Research Council Executive Agency. Neither the European Union nor the granting authority can be held responsible for them.

\end{acknowledgments}

\appendix

\section{Impact of a fully sampled IMF}
\label{app:a}

\begin{figure}[thbp]
    \includegraphics[width=1.\hsize]{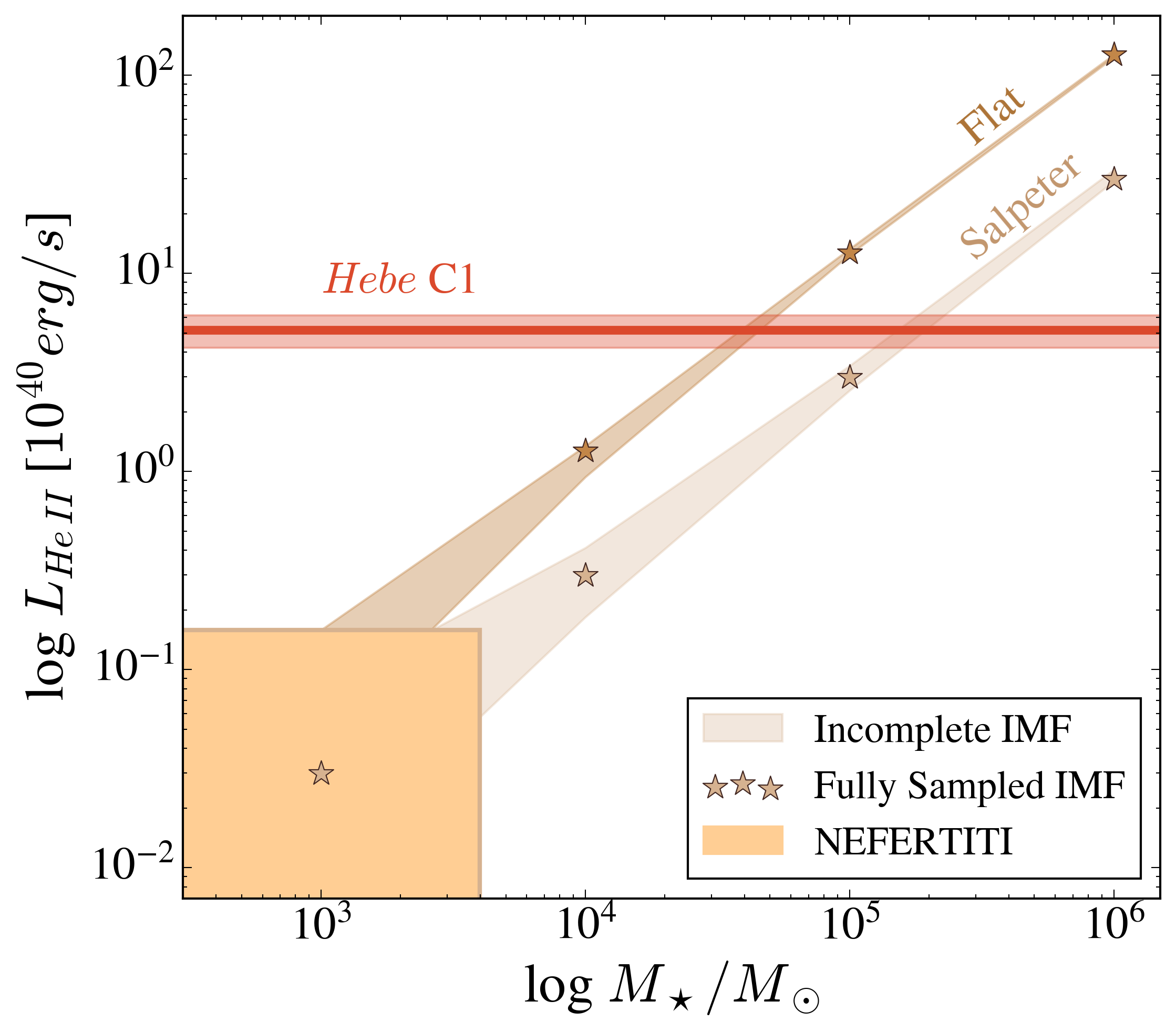} 
    \caption{Luminosity of HeII$\lambda1640$ versus stellar mass for a single burst of PopIII stars with an age of 1 Myr, forming according to PopIII IMF with different slopes: Flat (brown) or Salpeter ($x=2.35$, beige) with $M_{ch}=10 M_\odot$ (see eq.~\ref{e:Larson}). The stars represent a fully sampled IMF, while the shaded region shows the range obtained by randomly sampling the IMF. The orange region showcases the range of masses and luminosities of pure PopIII {\tt NEFERTITI} galaxies (IMF slope Salpeter, $M_{ch}=10 M_\odot$). The red line is the measured HeII of C1 in {\it Hebe} reported by \citealt{maiolino+26}.} 
  \label{fig:fig5}
\end{figure}

To assess the impact of assuming a fully sampled PopIII IMF in our parametric study, we perform 30 test models with the same random sampling technique implemented in {\tt NEFERTITI}. Consistent with Sec.~\ref{sec:IMF}, we consider PopIII star formation episodes producing total stellar masses of $[10^4, 10^5, 10^6] M_\odot$, and randomly form stars according to a flat IMF or a Salpeter slope ($x=2.35$) with $M_{ch}=10M_\odot$ (same assumed for Fig.~\ref{fig:fig1}). We then compute the resulting HeII luminosity by summing the contributions of individual PopIII stars, all assumed to have an age of 1 Myr, with fixed $\rm logU=-1$. 

Fig.~\ref{fig:fig5} compares the $\rm L_{HeII}$ resulting from these randomly sampled models with those employed in our Sec.~\ref{sec:IMF}, which instead assume fully sampled PopIII IMFs. 
We also report the predictions for pure PopIII {\tt NEFERTITI} galaxies of Fig.~\ref{fig:fig1}, which appear at $M_\star \leq 4 \cdot10^3M_\odot$.

We find that for large masses of PopIII stars formed ($M_\star \approx 10^6 M_\odot$), the differences in $\rm L_{HeII}$ are negligible, as the PopIII IMF is effectively fully sampled \citep[e.g., see][]{rossi2021ultra}. Conversely, discrepancies increase toward lower stellar masses, although the $\rm L_{HeII}$ range spanned by randomly sampled models remains centered around the fully sampled case. For example, at $M_\star=10^4\,M_\odot$, the fully sampled IMF with Salpeter slope yields $\rm L_{HeII}= 3 \cdot 10^{39} erg/s$, while incomplete sampling gives $\rm L_{HeII}=(1.8-4.2) \cdot 10^{39} erg/s$. 

In the lowest mass regime populated by 80 pure PopIII galaxies from {\tt NEFERTITI}, the scatter becomes significant, showcasing the importance of stochastic IMF sampling when modeling such very low-mass systems.

Finally, Fig.~\ref{fig:fig5} shows that a PopIII stellar mass of $M_\star = 4 \cdot 10^4\,M_\odot$ ($M_\star = 1.8 \cdot10^5\,M_\odot$) is required to match the observed $\rm HeII$ luminosity of the C1 component in {\it Hebe} under a Salpeter IMF (or a flat IMF). This is fully consistent with the results of Sec.~\ref{sec:IMF} (see in particular Fig.~\ref{fig:fig3}, top-right panel).

\bibliography{refer, codes}

@ARTICLE{jeong2026,
       author = {{Jeong}, Tae Bong and {Venditti}, Alessandra and {Bromm}, Volker and {Jeon}, Myoungwon and {Hsiao}, Tiger Yu-Yang and {Finkelstein}, Steven L. and {Chisholm}, John},
        title = "{How Massive Can a Population III Starburst Be? Simulating the First Galaxies with High Lyman-Werner Background}",
      journal = {arXiv e-prints},
         year = 2026,
        month = mar,
          eid = {arXiv:2603.23209},
        pages = {arXiv:2603.23209},
          doi = {10.48550/arXiv.2603.23209},
archivePrefix = {arXiv},
       eprint = {2603.23209},
 primaryClass = {astro-ph.GA},
       adsurl = {https://ui.adsabs.harvard.edu/abs/2026arXiv260323209J}
}

@ARTICLE{ubler+26,
       author = {{{\"U}bler}, Hannah and {Maiolino}, Roberto and {P{\'e}rez-Gonz{\'a}lez}, Pablo G. and {Isobe}, Yuki and {Jones}, Gareth C. and {Kumari}, Nimisha and {Charlot}, St{\'e}phane and {Rusta}, Elka and {Salvadori}, Stefania and {Nakajima}, Kimihiko and {Perna}, Michele and {Arribas}, Santiago and {Bunker}, Andrew J. and {Carniani}, Stefano and {D'Eugenio}, Francesco and {Rodr{\'\i}guez Del Pino}, Bruno and {Bertola}, Elena and {B{\"o}ker}, Torsten and {Chevallard}, Jacopo and {Circosta}, Chiara and {Cresci}, Giovanni and {Curti}, Mirko and {Curtis-Lake}, Emma and {Eisenstein}, Daniel J. and {Hainline}, Kevin and {Johnson}, Benjamin D. and {Parlanti}, Eleonora and {Rinaldi}, Pierluigi and {Robertson}, Brant and {Scholtz}, Jan and {Tacchella}, Sandro and {Venturi}, Giacomo and {Witstok}, Joris and {Zamora}, Sandra},
        title = "{GA-NIFS \& JADES: Confirmation of pristine gas near GN-z11}",
      journal = {arXiv e-prints},
         year = 2026,
        month = mar,
          eid = {arXiv:2603.20360},
        pages = {arXiv:2603.20360},
          doi = {10.48550/arXiv.2603.20360},
archivePrefix = {arXiv},
       eprint = {2603.20360},
 primaryClass = {astro-ph.GA},
       adsurl = {https://ui.adsabs.harvard.edu/abs/2026arXiv260320360U}
}

@ARTICLE{maiolino+26,
       author = {{Maiolino}, Roberto and {{\"U}bler}, Hannah and {Perna}, Michele and {Witstok}, Joris and {Jones}, Gareth C. and {Perez-Gonzalez}, Pablo G. and {Nakajima}, Kimihiko and {Rusta}, Elka and {Salvadori}, Stefania and {Tacchella}, Sandro and {Madau}, Piero and {Trussler}, James A.~A. and {D'Eugenio}, Francesco and {Ji}, Xihan and {Scholtz}, Jan and {Carniani}, Stefano and {Isobe}, Yuki and {Katz}, Harley and {Arribas}, Santiago and {Baker}, William M. and {B{\"o}ker}, Torsten and {Bromm}, Volker and {Bunker}, Andrew J. and {Charlot}, Stephane and {Chevallard}, Jacopo and {Curti}, Mirko and {Curtis-Lake}, Emma and {Eisenstein}, Daniel and {Egami}, Eiichi and {Ferrara}, Andrea and {Graziani}, Luca and {Hainline}, Kevin and {Helton}, Jakob M. and {Ivey}, Lucy and {Jonson}, Benjamin and {Koller}, Maria and {Kumari}, Nimisha and {Marconi}, Alessandro and {Mazzolari}, Giovanni and {Laporte}, Nicolas and {Parlanti}, Eleonora and {Pascalau}, Robert and {Pentericci}, Laura and {Rinaldi}, Pierluigi and {Robertson}, Brant and {Rodr{\'\i}guez Del Pino}, Bruno and {Schneider}, Raffaella and {Venditti}, Alessandra and {Venturi}, Giacomo and {Willmer}, Christopher N.~A. and {Witten}, Callum and {Zamora}, Sandra},
        title = "{The search for Population III: Confirmation of a HeII emitter with no metal lines at z=10.6}",
      journal = {arXiv e-prints},
         year = 2026,
        month = mar,
          eid = {arXiv:2603.20362},
        pages = {arXiv:2603.20362},
          doi = {10.48550/arXiv.2603.20362},
archivePrefix = {arXiv},
       eprint = {2603.20362},
 primaryClass = {astro-ph.GA},
       adsurl = {https://ui.adsabs.harvard.edu/abs/2026arXiv260320362M}
}

@ARTICLE{peng01,
       author = {{Oh}, S. Peng and {Haiman}, Zolt{\'a}n and {Rees}, Martin J.},
        title = "{HE II Recombination Lines from the First Luminous Objects}",
      journal = {\apj},
         year = 2001,
        month = may,
       volume = {553},
       number = {1},
        pages = {73-77},
          doi = {10.1086/320650},
archivePrefix = {arXiv},
       eprint = {astro-ph/0007351},
 primaryClass = {astro-ph},
       adsurl = {https://ui.adsabs.harvard.edu/abs/2001ApJ...553...73O}
}

@ARTICLE{Sun2026,
       author = {{Sun}, Fengwu and {Eisenstein}, Daniel J. and {D'Eugenio}, Francesco and {Hainline}, Kevin and {Helton}, Jakob M. and {Johnson}, Benjamin D. and {Lin}, Xiaojing and {Rieke}, Marcia and {Robertson}, Brant and {Tacchella}, Sandro and {Bunker}, Andrew J. and {Chevallard}, Jacopo and {Curtis-Lake}, Emma and {Egami}, Eiichi and {Hausen}, Ryan and {Ji}, Zhiyuan and {Lyu}, Jianwei and {Maiolino}, Roberto and {Rinaldi}, Pierluigi and {Sun}, Yang and {Trussler}, James A.~A. and {Williams}, Christina C. and {Willmer}, Christopher N.~A. and {Witstok}, Joris and {Wu}, Zihao and {Zhu}, Yongda},
        title = "{JADES: Discovery of Large Reservoirs of Small Dust Grains in the Circumgalactic Medium of Massive Galaxies at $z\sim3.5$ through Deep JWST/NIRCam Imaging and Grism Spectroscopy}",
      journal = {arXiv e-prints},
         year = 2026,
        month = jan,
          eid = {arXiv:2601.15961},
        pages = {arXiv:2601.15961},
          doi = {10.48550/arXiv.2601.15961},
archivePrefix = {arXiv},
       eprint = {2601.15961},
 primaryClass = {astro-ph.GA},
       adsurl = {https://ui.adsabs.harvard.edu/abs/2026arXiv260115961S}
}

@ARTICLE{Markov2025,
       author = {{Markov}, Vladan and {Gallerani}, Simona and {Ferrara}, Andrea and {Pallottini}, Andrea and {Parlanti}, Eleonora and {Mascia}, Fabio Di and {Sommovigo}, Laura and {Kohandel}, Mahsa},
        title = "{The evolution of dust attenuation in z {\ensuremath{\approx}} 2-12 galaxies observed by JWST}",
      journal = {Nature Astronomy},
         year = 2025,
        month = mar,
       volume = {9},
        pages = {458-468},
          doi = {10.1038/s41550-024-02426-1},
archivePrefix = {arXiv},
       eprint = {2402.05996},
 primaryClass = {astro-ph.GA},
       adsurl = {https://ui.adsabs.harvard.edu/abs/2025NatAs...9..458M}
}

@ARTICLE{Koutsouridou2025,
       author = {{Koutsouridou}, I. and {Sk{\'u}lad{\'o}ttir}, {\'A}. and {Salvadori}, S.},
      journal = {\aap},
         year = 2025,
        month = jul,
       volume = {699},
          eid = {A32},
        pages = {A32},
          doi = {10.1051/0004-6361/202554228},
archivePrefix = {arXiv},
       eprint = {2505.13607},
 primaryClass = {astro-ph.GA},
       adsurl = {https://ui.adsabs.harvard.edu/abs/2025A&A...699A..32K}
}

@ARTICLE{Storck2025,
       author = {{Storck}, Anatole and {Katz}, Harley and {Devriendt}, Julien and {Slyz}, Adrianne and {Cadiou}, Corentin and {Choustikov}, Nicholas and {Rey}, Martin P. and {Saxena}, Aayush and {Agertz}, Oscar and {Kimm}, Taysun},
        title = "{MEGATRON: The environments of Population III stars at Cosmic Dawn and their connection to present day galaxies}",
      journal = {arXiv e-prints},
         year = 2025,
        month = oct,
          eid = {arXiv:2510.06853},
        pages = {arXiv:2510.06853},
          doi = {10.48550/arXiv.2510.06853},
archivePrefix = {arXiv},
       eprint = {2510.06853},
 primaryClass = {astro-ph.GA},
       adsurl = {https://ui.adsabs.harvard.edu/abs/2025arXiv251006853S}
}

@ARTICLE{ventura2024,
       author = {{Ventura}, Emanuele M. and {Qin}, Yuxiang and {Balu}, Sreedhar and {Wyithe}, J. Stuart B.},
        title = "{Semi-analytic modelling of Pop. III star formation and metallicity evolution - I. Impact on the UV luminosity functions at z = 9-16}",
      journal = {\mnras},
         year = 2024,
        month = mar,
       volume = {529},
       number = {1},
        pages = {628-646},
          doi = {10.1093/mnras/stae567},
archivePrefix = {arXiv},
       eprint = {2401.07396},
 primaryClass = {astro-ph.GA},
       adsurl = {https://ui.adsabs.harvard.edu/abs/2024MNRAS.529..628V}
}

@ARTICLE{Hazlett2025,
       author = {{Hazlett}, Ryan and {Mead}, Jennifer and {Visbal}, Eli and {Bryan}, Greg L. and {Mac Low}, Mordecai-Mark and {Kulkarni}, Mihir and {Andersson}, Eric P. and {Brauer}, Kaley and {Wise}, John H.},
        title = "{From Primordial Stars to Early Galaxies: A Semi-Analytic Model Calibrated with Aeos and Renaissance}",
      journal = {arXiv e-prints},
         year = 2025,
        month = oct,
          eid = {arXiv:2510.11629},
        pages = {arXiv:2510.11629},
          doi = {10.48550/arXiv.2510.11629},
archivePrefix = {arXiv},
       eprint = {2510.11629},
 primaryClass = {astro-ph.GA},
       adsurl = {https://ui.adsabs.harvard.edu/abs/2025arXiv251011629H}
}

@ARTICLE{Jaacks2019,
       author = {{Jaacks}, Jason and {Finkelstein}, Steven L. and {Bromm}, Volker},
        title = "{Legacy of star formation in the pre-reionization universe}",
      journal = {\mnras},
         year = 2019,
        month = sep,
       volume = {488},
       number = {2},
        pages = {2202-2221},
          doi = {10.1093/mnras/stz1529},
archivePrefix = {arXiv},
       eprint = {1804.07372},
 primaryClass = {astro-ph.GA},
       adsurl = {https://ui.adsabs.harvard.edu/abs/2019MNRAS.488.2202J}
}

@ARTICLE{Schneider2004,
       author = {{Schneider}, R. and {Ferrara}, A. and {Salvaterra}, R.},
        title = "{Dust formation in very massive primordial supernovae}",
      journal = {\mnras},
         year = 2004,
        month = jul,
       volume = {351},
       number = {4},
        pages = {1379-1386},
          doi = {10.1111/j.1365-2966.2004.07876.x},
archivePrefix = {arXiv},
       eprint = {astro-ph/0307087},
 primaryClass = {astro-ph},
       adsurl = {https://ui.adsabs.harvard.edu/abs/2004MNRAS.351.1379S}
}

@ARTICLE{Nozawa2003,
       author = {{Nozawa}, Takaya and {Kozasa}, Takashi and {Umeda}, Hideyuki and {Maeda}, Keiichi and {Nomoto}, Ken'ichi},
        title = "{Dust in the Early Universe: Dust Formation in the Ejecta of Population III Supernovae}",
      journal = {\apj},
         year = 2003,
        month = dec,
       volume = {598},
       number = {2},
        pages = {785-803},
          doi = {10.1086/379011},
archivePrefix = {arXiv},
       eprint = {astro-ph/0307108},
 primaryClass = {astro-ph},
       adsurl = {https://ui.adsabs.harvard.edu/abs/2003ApJ...598..785N}
}

@ARTICLE{gandolfi+2026,
       author = {{Gandolfi}, G. and {Rodighiero}, G. and {Castellano}, M. and {Fontana}, A. and {Santini}, P. and {Dickinson}, M. and {Finkelstein}, S. and {Catone}, M. and {Calabr{\`o}}, A. and {Merlin}, E. and {Pentericci}, L. and {Bisigello}, L. and {Grazian}, A. and {Napolitano}, L. and {Vulcani}, B. and {Taylor}, A.~J. and {Arrabal Haro}, P. and {Kirkpatrick}, A. and {Backhaus}, B.~E. and {Holwerda}, B.~W. and {Giulietti}, M. and {Bianchetti}, A. and {Cassata}, P. and {Cleri}, N.~J. and {Daddi}, E. and {Ferguson}, H.~C. and {Girardi}, G. and {Hirschmann}, M. and {Koekemoer}, A.~M. and {Lapi}, A. and {Pacucci}, F. and {P{\'e}rez-Gonz{\'a}lez}, P.~G. and {de la Vega}, A. and {Vietri}, A. and {Wilkins}, S. and {Yung}, L.~Y.~A. and {Bagley}, M. and {Bhatawdekar}, R. and {Kartaltepe}, J. and {Papovich}, C. and {Pirzkal}, N.},
        title = "{Mysteries of Capotauro: Investigating the puzzling nature of an extreme F356W-dropout}",
      journal = {\aap},
         year = 2026,
        month = feb,
       volume = {706},
          eid = {A364},
        pages = {A364},
          doi = {10.1051/0004-6361/202557061},
archivePrefix = {arXiv},
       eprint = {2509.01664},
 primaryClass = {astro-ph.GA},
       adsurl = {https://ui.adsabs.harvard.edu/abs/2026A&A...706A.364G}
}

@ARTICLE{ferrara+2026,
       author = {{Ferrara}, Andrea and {Carniani}, Stefano and {Morishita}, Takahiro and {Stiavelli}, Massimo},
        title = "{Possible evidence for a pair-instability supernova nature of ultra-early JWST sources}",
      journal = {arXiv e-prints},
         year = 2026,
        month = jan,
          eid = {arXiv:2601.07374},
        pages = {arXiv:2601.07374},
          doi = {10.48550/arXiv.2601.07374},
archivePrefix = {arXiv},
       eprint = {2601.07374},
 primaryClass = {astro-ph.GA},
       adsurl = {https://ui.adsabs.harvard.edu/abs/2026arXiv260107374F}
}

@ARTICLE{morishita+25,
       author = {{Morishita}, Takahiro and {Liu}, Zhaoran and {Stiavelli}, Massimo and {Treu}, Tommaso and {Bergamini}, Pietro and {Zhang}, Yechi},
        title = "{Pristine Massive Star Formation Caught at the Break of Cosmic Dawn}",
      journal = {arXiv e-prints},
         year = 2025,
        month = jul,
          eid = {arXiv:2507.10521},
        pages = {arXiv:2507.10521},
          doi = {10.48550/arXiv.2507.10521},
archivePrefix = {arXiv},
       eprint = {2507.10521},
 primaryClass = {astro-ph.CO},
       adsurl = {https://ui.adsabs.harvard.edu/abs/2025arXiv250710521M}
}

@ARTICLE{vanzella+26,
       author = {{Vanzella}, E. and {Messa}, M. and {Zanella}, A. and {Bolamperti}, A. and {Castellano}, M. and {Loiacono}, F. and {Bergamini}, P. and {Roberts Borsani}, G. and {Adamo}, A. and {Fontana}, A. and {Treu}, T. and {Calura}, F. and {Grillo}, C. and {Lombardi}, M. and {Rosati}, P. and {Gilli}, R. and {Meneghetti}, M.},
        title = "{A pristine, star-forming complex at z = 4.19}",
      journal = {\aap},
         year = 2026,
        month = jan,
       volume = {705},
          eid = {L12},
        pages = {L12},
          doi = {10.1051/0004-6361/202557153},
archivePrefix = {arXiv},
       eprint = {2509.07073},
 primaryClass = {astro-ph.GA},
       adsurl = {https://ui.adsabs.harvard.edu/abs/2026A&A...705L..12V}
}

@ARTICLE{rusta+25,
       author = {{Rusta}, Elka and {Salvadori}, Stefania and {Gelli}, Viola and {Schaerer}, Daniel and {Marconi}, Alessandro and {Koutsouridou}, Ioanna and {Carniani}, Stefano},
        title = "{Metal-polluted Population III Galaxies and How to Find Them}",
      journal = {\apjl},
         year = 2025,
        month = aug,
       volume = {989},
       number = {2},
          eid = {L32},
        pages = {L32},
          doi = {10.3847/2041-8213/adf4e3},
archivePrefix = {arXiv},
       eprint = {2506.17400},
 primaryClass = {astro-ph.GA},
       adsurl = {https://ui.adsabs.harvard.edu/abs/2025ApJ...989L..32R}
}

@ARTICLE{Nakajima+25,
       author = {{Nakajima}, Kimihiko and {Ouchi}, Masami and {Harikane}, Yuichi and {Vanzella}, Eros and {Ono}, Yoshiaki and {Isobe}, Yuki and {Nishigaki}, Moka and {Tsujimoto}, Takuji and {Nakamura}, Fumitaka and {Xu}, Yi and {Umeda}, Hiroya and {Zhang}, Yechi},
        title = "{An Ultra-Faint, Chemically Primitive Galaxy Forming at the Epoch of Reionization}",
      journal = {arXiv e-prints},
         year = 2025,
        month = jun,
          eid = {arXiv:2506.11846},
        pages = {arXiv:2506.11846},
          doi = {10.48550/arXiv.2506.11846},
archivePrefix = {arXiv},
       eprint = {2506.11846},
 primaryClass = {astro-ph.GA},
       adsurl = {https://ui.adsabs.harvard.edu/abs/2025arXiv250611846N}
}

@ARTICLE{Vanni2024,
       author = {{Vanni}, Irene and {Salvadori}, Stefania and {D'Odorico}, Valentina and {Becker}, George D. and {Cupani}, Guido},
        title = "{Chemical Diagnostics to Unveil Environments Enriched by First Stars}",
      journal = {\apjl},
         year = 2024,
        month = jun,
       volume = {967},
       number = {2},
          eid = {L22},
        pages = {L22},
          doi = {10.3847/2041-8213/ad46fa},
archivePrefix = {arXiv},
       eprint = {2402.18640},
 primaryClass = {astro-ph.GA},
       adsurl = {https://ui.adsabs.harvard.edu/abs/2024ApJ...967L..22V}
}

@ARTICLE{Koustouridou2024,
       author = {{Koutsouridou}, Ioanna and {Salvadori}, Stefania and {Sk{\'u}lad{\'o}ttir}, {\'A}sa},
        title = "{True Pair-instability Supernova Descendant: Implications for the First Stars' Mass Distribution}",
      journal = {\apjl},
         year = 2024,
        month = feb,
       volume = {962},
       number = {2},
          eid = {L26},
        pages = {L26},
          doi = {10.3847/2041-8213/ad2466},
       adsurl = {https://ui.adsabs.harvard.edu/abs/2024ApJ...962L..26K}
}

@ARTICLE{Koutsouridou2023,
       author = {{Koutsouridou}, I. and {Salvadori}, S. and {Sk{\'u}lad{\'o}ttir}, {\'A}. and {Rossi}, M. and {Vanni}, I. and {Pagnini}, G.},
        title = "{The energy distribution of the first supernovae}",
      journal = {\mnras},
         year = 2023,
        month = oct,
       volume = {525},
       number = {1},
        pages = {190-210},
          doi = {10.1093/mnras/stad2304},
archivePrefix = {arXiv},
       eprint = {2309.00045},
 primaryClass = {astro-ph.GA},
       adsurl = {https://ui.adsabs.harvard.edu/abs/2023MNRAS.525..190K}
}

@ARTICLE{Schaerer2002,
       author = {{Schaerer}, D.},
        title = "{On the properties of massive Population III stars and metal-free stellar populations}",
      journal = {\aap},
         year = 2002,
        month = jan,
       volume = {382},
        pages = {28-42},
          doi = {10.1051/0004-6361:20011619},
archivePrefix = {arXiv},
       eprint = {astro-ph/0110697},
 primaryClass = {astro-ph},
       adsurl = {https://ui.adsabs.harvard.edu/abs/2002A&A...382...28S}
}

@article{larson1998early,
  title={Early star formation and the evolution of the stellar initial mass function in galaxies},
  author={Larson, Richard B},
  journal={Monthly Notices of the Royal Astronomical Society},
  volume={301},
  number={2},
  pages={569--581},
  year={1998},
  publisher={Wiley Online Library}
}

@ARTICLE{Bromm2013,
       author = {{Bromm}, Volker},
        title = "{Formation of the first stars}",
      journal = {Reports on Progress in Physics},
         year = 2013,
        month = nov,
       volume = {76},
       number = {11},
          eid = {112901},
        pages = {112901},
          doi = {10.1088/0034-4885/76/11/112901},
archivePrefix = {arXiv},
       eprint = {1305.5178},
 primaryClass = {astro-ph.CO},
       adsurl = {https://ui.adsabs.harvard.edu/abs/2013RPPh...76k2901B}
}

@ARTICLE{Venditti2023,
       author = {{Venditti}, Alessandra and {Graziani}, Luca and {Schneider}, Raffaella and {Pentericci}, Laura and {Di Cesare}, Claudia and {Maio}, Umberto and {Omukai}, Kazuyuki},
        title = "{A needle in a haystack? Catching Pop\textbackslash III stars in the Epoch of Reionization: I. Pop\textbackslash III star forming environments}",
      journal = {arXiv e-prints},
         year = 2023,
        month = jan,
          eid = {arXiv:2301.10259},
        pages = {arXiv:2301.10259},
          doi = {10.48550/arXiv.2301.10259},
archivePrefix = {arXiv},
       eprint = {2301.10259},
 primaryClass = {astro-ph.GA},
       adsurl = {https://ui.adsabs.harvard.edu/abs/2023arXiv230110259V}
}

@article{rossi2021ultra,
  title={Ultra-faint dwarf galaxies: unveiling the minimum mass of the first stars},
  author={Rossi, Martina and Salvadori, Stefania and Sk{\'u}lad{\'o}ttir, {\'A}sa},
  journal={Monthly Notices of the Royal Astronomical Society},
  year={2021}
}

@article{salvadori2019probing,
  title={Probing the existence of very massive first stars},
  author={Salvadori, S and Bonifacio, P and Caffau, E and Korotin, S and Andreevsky, S and Spite, M and Sk{\'u}lad{\'o}ttir, {\'A}},
  journal={Monthly Notices of the Royal Astronomical Society},
  volume={487},
  number={3},
  pages={4261--4284},
  year={2019},
  publisher={Oxford University Press}
}

@article{zackrisson+11,
       author = {{Zackrisson}, Erik and {Rydberg}, Claes-Erik and {Schaerer}, Daniel and {{\"O}stlin}, G{\"o}ran and {Tuli}, Manan},
        title = "{The Spectral Evolution of the First Galaxies. I. James Webb Space Telescope Detection Limits and Color Criteria for Population III Galaxies}",
      journal = {\apj},
         year = 2011,
        month = oct,
       volume = {740},
       number = {1},
          eid = {13},
        pages = {13},
          doi = {10.1088/0004-637X/740/1/13},
archivePrefix = {arXiv},
       eprint = {1105.0921},
 primaryClass = {astro-ph.CO},
       adsurl = {https://ui.adsabs.harvard.edu/abs/2011ApJ...740...13Z}
}

@ARTICLE{Raiter+10,
       author = {{Raiter}, A. and {Schaerer}, D. and {Fosbury}, R.~A.~E.},
        title = "{Predicted UV properties of very metal-poor starburst galaxies}",
      journal = {\aap},
         year = 2010,
        month = nov,
       volume = {523},
          eid = {A64},
        pages = {A64},
          doi = {10.1051/0004-6361/201015236},
archivePrefix = {arXiv},
       eprint = {1008.2114},
 primaryClass = {astro-ph.CO},
       adsurl = {https://ui.adsabs.harvard.edu/abs/2010A&A...523A..64R}
}

@ARTICLE{rusta+24,
       author = {{Rusta}, Elka and {Salvadori}, Stefania and {Gelli}, Viola and {Koutsouridou}, Ioanna and {Marconi}, Alessandro},
        title = "{Linking High-z and Low-z: Are We Observing the Progenitors of the Milky Way with JWST?}",
      journal = {\apjl},
         year = 2024,
        month = oct,
       volume = {974},
       number = {2},
          eid = {L35},
        pages = {L35},
          doi = {10.3847/2041-8213/ad833d},
archivePrefix = {arXiv},
       eprint = {2407.06255},
 primaryClass = {astro-ph.GA},
       adsurl = {https://ui.adsabs.harvard.edu/abs/2024ApJ...974L..35R}
}

@ARTICLE{schaerer+03,
       author = {{Schaerer}, D.},
        title = "{The transition from Population III to normal galaxies: Lyalpha and He II emission and the ionising properties of high redshift starburst galaxies}",
      journal = {\aap},
         year = 2003,
        month = jan,
       volume = {397},
        pages = {527-538},
          doi = {10.1051/0004-6361:20021525},
archivePrefix = {arXiv},
       eprint = {astro-ph/0210462},
 primaryClass = {astro-ph},
       adsurl = {https://ui.adsabs.harvard.edu/abs/2003A&A...397..527S}
}

@ARTICLE{chatzikos+23,
       author = {{Chatzikos}, M. and {Bianchi}, S. and {Camilloni}, F. and {Chakraborty}, P. and {Gunasekera}, C.~M. and {Guzm{\'a}n}, F. and {Milby}, J.~S. and {Sarkar}, A. and {Shaw}, G. and {van Hoof}, P.~A.~M. and {Ferland}, G.~J.},
        title = "{The 2023 Release of Cloudy}",
      journal = {\rmxaa},
         year = 2023,
        month = oct,
       volume = {59},
        pages = {327-343},
          doi = {10.22201/ia.01851101p.2023.59.02.12},
archivePrefix = {arXiv},
       eprint = {2308.06396},
 primaryClass = {astro-ph.GA},
       adsurl = {https://ui.adsabs.harvard.edu/abs/2023RMxAA..59..327C}
}

@ARTICLE{ferland+98,
       author = {{Ferland}, G.~J. and {Korista}, K.~T. and {Verner}, D.~A. and {Ferguson}, J.~W. and {Kingdon}, J.~B. and {Verner}, E.~M.},
        title = "{CLOUDY 90: Numerical Simulation of Plasmas and Their Spectra}",
      journal = {\pasp},
         year = 1998,
        month = jul,
       volume = {110},
       number = {749},
        pages = {761-778},
          doi = {10.1086/316190},
       adsurl = {https://ui.adsabs.harvard.edu/abs/1998PASP..110..761F}
}

@ARTICLE{maiolino+24,
       author = {{Maiolino}, Roberto and {{\"U}bler}, Hannah and {Perna}, Michele and {Scholtz}, Jan and {D'Eugenio}, Francesco and {Witten}, Callum and {Laporte}, Nicolas and {Witstok}, Joris and {Carniani}, Stefano and {Tacchella}, Sandro and {Baker}, William M. and {Arribas}, Santiago and {Nakajima}, Kimihiko and {Eisenstein}, Daniel J. and {Bunker}, Andrew J. and {Charlot}, St{\'e}phane and {Cresci}, Giovanni and {Curti}, Mirko and {Curtis-Lake}, Emma and {de Graaff}, Anna and {Egami}, Eiichi and {Ji}, Zhiyuan and {Johnson}, Benjamin D. and {Kumari}, Nimisha and {Looser}, Tobias J. and {Maseda}, Michael and {Nelson}, Erica and {Robertson}, Brant and {Rodr{\'\i}guez Del Pino}, Bruno and {Sandles}, Lester and {Simmonds}, Charlotte and {Smit}, Renske and {Sun}, Fengwu and {Venturi}, Giacomo and {Williams}, Christina C. and {Willmer}, Christopher N.~A.},
        title = "{JADES. Possible Population III signatures at z = 10.6 in the halo of GN-z11}",
      journal = {\aap},
         year = 2024,
        month = jul,
       volume = {687},
          eid = {A67},
        pages = {A67},
          doi = {10.1051/0004-6361/202347087},
archivePrefix = {arXiv},
       eprint = {2306.00953},
 primaryClass = {astro-ph.GA},
       adsurl = {https://ui.adsabs.harvard.edu/abs/2024A&A...687A..67M}
}

@ARTICLE{vanzella+23,
       author = {{Vanzella}, E. and {Loiacono}, F. and {Bergamini}, P. and {Me{\v{s}}tri{\'c}}, U. and {Castellano}, M. and {Rosati}, P. and {Meneghetti}, M. and {Grillo}, C. and {Calura}, F. and {Mignoli}, M. and {Brada{\v{c}}}, M. and {Adamo}, A. and {Rihtar{\v{s}}i{\v{c}}}, G. and {Dickinson}, M. and {Gronke}, M. and {Zanella}, A. and {Annibali}, F. and {Willott}, C. and {Messa}, M. and {Sani}, E. and {Acebron}, A. and {Bolamperti}, A. and {Comastri}, A. and {Gilli}, R. and {Caputi}, K.~I. and {Ricotti}, M. and {Gruppioni}, C. and {Ravindranath}, S. and {Mercurio}, A. and {Strait}, V. and {Martis}, N. and {Pascale}, R. and {Caminha}, G.~B. and {Annunziatella}, M. and {Nonino}, M.},
        title = "{An extremely metal-poor star complex in the reionization era: Approaching Population III stars with JWST}",
      journal = {\aap},
         year = 2023,
        month = oct,
       volume = {678},
          eid = {A173},
        pages = {A173},
          doi = {10.1051/0004-6361/202346981},
archivePrefix = {arXiv},
       eprint = {2305.14413},
 primaryClass = {astro-ph.GA},
       adsurl = {https://ui.adsabs.harvard.edu/abs/2023A&A...678A.173V}
}

@ARTICLE{tumlinson+00,
       author = {{Tumlinson}, Jason and {Shull}, J. Michael},
        title = "{Zero-Metallicity Stars and the Effects of the First Stars on Reionization}",
      journal = {\apjl},
         year = 2000,
        month = jan,
       volume = {528},
       number = {2},
        pages = {L65-L68},
          doi = {10.1086/312432},
archivePrefix = {arXiv},
       eprint = {astro-ph/9911339},
 primaryClass = {astro-ph},
       adsurl = {https://ui.adsabs.harvard.edu/abs/2000ApJ...528L..65T}
}

@ARTICLE{Nakajima+22,
       author = {{Nakajima}, K. and {Maiolino}, R.},
        title = "{Diagnostics for PopIII galaxies and direct collapse black holes in the early universe}",
      journal = {\mnras},
         year = 2022,
        month = jul,
       volume = {513},
       number = {4},
        pages = {5134-5147},
          doi = {10.1093/mnras/stac1242},
archivePrefix = {arXiv},
       eprint = {2204.11870},
 primaryClass = {astro-ph.GA},
       adsurl = {https://ui.adsabs.harvard.edu/abs/2022MNRAS.513.5134N}
}

@ARTICLE{klessen+23,
       author = {{Klessen}, Ralf S. and {Glover}, Simon C.~O.},
        title = "{The First Stars: Formation, Properties, and Impact}",
      journal = {\araa},
         year = 2023,
        month = aug,
       volume = {61},
        pages = {65-130},
          doi = {10.1146/annurev-astro-071221-053453},
archivePrefix = {arXiv},
       eprint = {2303.12500},
 primaryClass = {astro-ph.CO},
       adsurl = {https://ui.adsabs.harvard.edu/abs/2023ARA&A..61...65K}
}

@ARTICLE{inoue+11,
       author = {{Inoue}, Akio K.},
        title = "{Rest-frame ultraviolet-to-optical spectral characteristics of extremely metal-poor and metal-free galaxies}",
      journal = {\mnras},
         year = 2011,
        month = aug,
       volume = {415},
       number = {3},
        pages = {2920-2931},
          doi = {10.1111/j.1365-2966.2011.18906.x},
archivePrefix = {arXiv},
       eprint = {1102.5150},
 primaryClass = {astro-ph.CO},
       adsurl = {https://ui.adsabs.harvard.edu/abs/2011MNRAS.415.2920I}
}

@ARTICLE{lecroq+25,
       author = {{Lecroq}, Marie and {Charlot}, St{\'e}phane and {Bressan}, Alessandro and {Bruzual}, Gustavo and {Costa}, Guglielmo and {Iorio}, Giuliano and {Mapelli}, Michela and {Santoliquido}, Filippo and {Shepherd}, Kendall and {Spera}, Mario},
        title = "{A new prescription for the spectral properties of population III stellar populations}",
      journal = {\aap},
         year = 2025,
        month = mar,
       volume = {695},
          eid = {A17},
        pages = {A17},
          doi = {10.1051/0004-6361/202452463},
archivePrefix = {arXiv},
       eprint = {2502.14028},
 primaryClass = {astro-ph.GA},
       adsurl = {https://ui.adsabs.harvard.edu/abs/2025A&A...695A..17L}
}
\bibliographystyle{aasjournalv7}

\end{document}